\documentclass[prb,aps,groupedaddress,twocolumn]{revtex4}
\usepackage{graphicx}
\usepackage{amsmath}
\begin{document}
\title{Optimized Jastrow-Slater wave functions for ground and excited states: \\ 
Application to the lowest states of ethene}

\author{Friedemann Schautz} \author{Claudia Filippi}
\affiliation{Instituut-Lorentz, Universiteit Leiden, Niels Bohrweg 2, Leiden, NL-2333
CA, The Netherlands}
\date{\today}

\begin{abstract}
A quantum Monte Carlo method is presented for determining multi-determinantal
Jastrow-Slater wave functions for which the energy is stationary with respect to
the simultaneous optimization of orbitals and configuration interaction 
coefficients. The approach is within the framework of the so-called energy fluctuation 
potential method which minimizes the energy in an iterative fashion based on Monte
Carlo sampling and a fitting of the local energy fluctuations. The optimization of the 
orbitals is combined with the optimization of the configuration interaction coefficients 
through the use of additional single excitations to a set of external orbitals. A new 
set of orbitals is then obtained from the natural orbitals of this enlarged configuration 
interaction expansion. For excited states, the approach is extended  to treat the average 
of several states within the same irreducible representation of the pointgroup of 
the molecule.
The relationship of our optimization method with the stochastic reconfiguration 
technique by Sorella {\it et al.} is examined. Finally, the performance of our approach 
is illustrated with the lowest states of ethene, in particular with the difficult case of 
the $1{}^1\mbox{B}_{1u}$ state.
\end{abstract}

\maketitle

\section{\label{sec:introduction}Introduction}

Quantum Monte Carlo (QMC) methods have been successfully employed over the 
last decade to compute ground state electronic properties of large molecules and 
solids~\cite{review}.  Compared to other electronic structure approaches, QMC has the advantage 
that it can be applied to sufficiently large systems and still provide an accurate 
description of both dynamical and static electronic correlation.  Recently, QMC 
methods have also been used for the calculations of excited 
states~\cite{grossman01,needs01,excited_nano_qmc,porphyrin}, a rather 
novel direction where relatively little experience exists. 

In both variational Monte Carlo (VMC) and diffusion Monte Carlo (DMC), the many-body 
trial wave function determines the quality of the calculation and is commonly chosen 
to be of the Jastrow-Slater type: a single- or multi-determinant wave function is multiplied 
by a correlation Jastrow factor to partly account for dynamic electronic correlation. 
While the Jastrow factor is generally optimized within QMC using the variance 
minimization method~\cite{varmin}, the determinantal component of the wave function 
is computed with methods such as Hartree-Fock, multi-configuration self consistent 
field (MCSCF, CASSCF) or small scale configuration interaction (CI), and, in most 
cases, is left unchanged when the Jastrow factor is added.
However, the Jastrow factor is a positive function of the inter-particle coordinates 
and, therefore, does not change the nodal surface of the trial wave function. 
Consequently, the fixed-node DMC energy is solely determined by the {\em determinantal
part} of the trial wave function~\cite{non_loc_pseudo}. 
Any approach aimed at improving the fixed-node DMC energy using a Jastrow-Slater trial
wave function must therefore reoptimize the determinantal part in the presence
of the Jastrow factor.

Recently, the energy fluctuation potential (EFP) approach to the optimization of 
Jastrow-Slater wave functions has been developed and applied to ground state 
calculations: the energy has been minimized with respect to a subset of the orbital 
parameters in single-determinant wave functions~\cite{efpcs}, the CI coefficients in 
multi-determinant wave functions~\cite{efpfs}, and the Jastrow parameters in periodic 
systems~\cite{efpds}. In the EFP method, the optimization is performed iteratively
through Monte Carlo sampling and the fitting of the energy fluctuations.
Here, we extend the EFP method to the simultaneous full optimization of CI 
coefficients and single-particle orbitals by reformulating the optimization of the 
orbitals as a super-CI approach~\cite{ruedenberg} for Jastrow-Slater wave function: 
the orbital variations are expressed as single excitations to a set of external 
orbitals and the improved orbitals are then obtained from the natural orbitals of this
super-CI expansion. 
Moreover, for excited states, we further develop the EFP technique to treat an 
average of several states within the same irreducible representation of the 
pointgroup of the molecule. With these additions, the method closely resembles  
the state-averaged MCSCF approach~\cite{samcscf}. When one treats multiple
states, the resulting Jastrow-Slater wave functions are however not orthogonal 
since orthogonality is only ensured for the determinantal parts. 

To illustrate the performance of our optimization approach, we study the 
$1{}^1\mbox{B}_{1u}$ state of ethene which has been the subject of much
theoretical debate and reconsideration over more than twenty 
years~\cite{graham,wiberg,foresman,roos,malrieu,gemein,buenker,davidson,lischka,grimme,koch}.
The difficulty in describing the $1{}^1\mbox{B}_{1u}$ can be mainly ascribed to the 
fact that any method insufficiently accounting for electron correlation will 
strongly mix this valence state with Rydberg states.
Explicit inclusion of dynamic correlation in the reference wave function 
appears to be necessary to avoid such mixing but this renders the construction 
of a reference for highly correlated quantum chemical calculations particularly 
complicated. In addition, the result is sensitive to the choice of the basis
and to the optimization of the occupied $\sigma$- orbitals which cannot all be treated as
frozen. QMC calculations for the $1{}^1\mbox{B}_{1u}$ state are also affected 
by similar problems and, to obtain a satisfactory description of this state, we find 
it necessary to optimize the determinantal part of the wave function in the presence of 
the Jastrow factor, that is to include the feedback of dynamic correlation on the 
determinantal reference. The Rydberg character of the initial trial wave function
is successfully corrected by our EFP optimization, and
the DMC excitation energy is lowered by about 0.5~eV upon optimization of the orbitals
and brought in very good agreement with the most sophisticated quantum chemical 
results. Our optimization approach is very efficient and robust, converging in a 
small number of iterations even when the initial trial wave function is worsened by making
the single-particle basis more diffuse. We find that the EFP optimization has
superior convergence properties than the stochastic reconfiguration method~\cite{sr1,sr2} by
Sorella {\it et al.}
Finally, we consider two Rydberg-like states $2{}^1\mbox{A}_{g}$ and  
$2{}^1\mbox{B}_{1u}$ as examples of states which are not the lowest ones in their 
irreducible representations and for which we successfully employ our extension of the 
EFP approach to the simultaneous optimization of multiple states. 
In a related paper~\cite{excited_sbf}, we further investigate the state-average EFP
approach by studying conformational changes in formaldimine, formaldehyde and a 
protonated Schiff-base model, and find that the optimization of excited states 
in the absence of symmetry constraints always works reliably. 

In Sec.~\ref{sec:efp}, we review the EFP approach and introduce our method 
to fully optimize CI and orbital coefficients for ground and excited states. 
The relationship with the stochastic reconfiguration method by 
Sorella {\it et al.} is also examined. 
Computational details are given in Sec.~\ref{sec:comp} and the numerical results 
for ethene are shown in Sec.~\ref{sec:results}. Finally, we present our 
conclusions in Sec.~\ref{sec:conclusion}.

\section{\label{sec:efp}Optimization of Jastrow-Slater wave functions}

The trial wave functions commonly used in quantum Monte Carlo calculations are 
of the Jastrow-Slater form:
\begin{eqnarray}
\Psi_{\rm T} = {\cal J}\;\Phi_{\rm T} = {\cal J}\; \sum_i c_i C_i\,,
\label{wf}
\end{eqnarray}
where ${\cal J}$ is the Jastrow factor which explicitly depends on electron-electron
separations and partially accounts for dynamic correlation. The $C_i$'s are a set 
of spin-adapted configuration state functions (CSF) expressed as linear 
combinations of Slater determinants:
\begin{eqnarray}
C_i = \sum_j d_{ij} D_j\,.
\label{eq:csf}
\end{eqnarray}
For a given Jastrow factor, we want to minimize the energy with respect to the 
CI coefficients $c_i$ and the orbitals in the determinantal part of the wave
function.

\subsection{Optimization of CI coefficients}

Let us assume that the determinantal component of the wave function $\Phi_{\rm T}$ 
(Eq.~\ref{wf}) is an eigenstate  of a CI Hamiltonian in the basis of the configuration 
state functions $C_i$.  Since the configuration state functions $C_i$ and the orthonormal 
eigenstates $\Phi_i$ of this Hamiltonian span the same space, we can express the 
infinitesimal variations in the $c_i$ coefficients of $\Phi_{\rm T}$ as variations 
with respect to the eigenstates $\Phi_i$ other than $\Phi_{\rm T}$:
\begin{eqnarray}
\Psi_{\rm T}={\cal J}\Phi_0\rightarrow 
\Psi_{\rm T}'={\cal J}\left(\Phi_0+\sum_{k>0} \delta_k \Phi_k\right)\,,
\label{eq:delta}
\end{eqnarray}
where, without loss of generality, we have set $\Phi_{\rm T}$ equal to the lowest 
eigenstate $\Phi_0$.
The corresponding derivatives of the energy are given by:
\begin{eqnarray}
\left.\frac{\partial E}{\partial\,\delta_k}\right|_{\delta=0}&=&
\left.\frac{\partial}{\partial\delta_k}\frac{\langle\Psi_{\rm T}'|{\cal H}
|\Psi_{\rm T}'\rangle}{\langle\Psi_{\rm T}'|\Psi_{\rm T}'\rangle}
\right|_{\delta=0}\nonumber\\
&=&\langle(E_L-\bar{E})\,(O_k-\bar{O}_k)\rangle\,.
\label{eq:deriv}
\end{eqnarray}
where $\langle\cdot\rangle$ denotes the average with respect to the 
square of trial wave function, $|\Psi_{\rm T}|^2$, which can be conveniently computed by Monte 
Carlo sampling.
We defined $\bar{E}=\langle E_L\rangle$ and $\bar{O}_k=\langle O_k \rangle$ where 
\begin{eqnarray}
O_k=\frac{\Phi_k}{\Phi_0}\quad \mbox{and}\quad E_L=\frac{{\cal H}\Psi_T}{\Psi_T}.
\end{eqnarray}
The energy is stationary with respect to variations of the CI coefficients 
if these derivatives are zero. As shown in Eq.~\ref{eq:deriv}, 
the fluctuations of the local energy become then uncorrelated with the fluctuations 
of the functions $O_k$, which means that the remaining fluctuations of the
local energy cannot be further reduced 
by adding some combination of the functions $O_k$. 

The energy fluctuation potential (EFP) method is based on this last observation, 
and reformulates the energy minimization problem as a least-squares fit of the
fluctuations of the 
local energy with an arbitrary combination of the functions $O_k$: 
\begin{eqnarray}
\chi^2 = \langle (E_L - \sum_k V_k O_k )^2 \rangle \,.
\label{eq:chisq}
\end{eqnarray}
A minimization of $\chi^2$ with respect to the parameters 
$V_k$ leads to the following set of linear equations~\cite{directmin}:
\begin{eqnarray}
\langle E_L O_m\rangle = 
\sum_k V_k\langle O_k O_m \rangle\,.
\end{eqnarray}  
The equation for the trial state $\Phi_0$ is simply
\begin{eqnarray}
\bar{E} = V_0 + \sum_{k>0} V_k \langle O_k \rangle\,,
\end{eqnarray}
and can be used to eliminate $V_0$ from the other equations which, for
$m>0$, become:
\begin{eqnarray}
\langle \Delta E\Delta O_m \rangle=
\sum_{k>0} V_k \langle \Delta O_k \Delta O_m \rangle\,,
\label{eq:lineq_qmc}
\end{eqnarray}
with $\Delta E = E_L - \langle E_L \rangle$ and 
$\Delta O_m = O_m - \langle O_m \rangle$.
The left-hand side of these equations correspond to the derivatives of the
energy with respect to variations in the CI coefficients (Eq.~\ref{eq:deriv}).
Therefore, the fitting parameters $V_{k>0}$ are all zero if and only if 
all the derivatives of the energy are zero~\cite{jastrow_linear}.

For an arbitrary trial wave function $\Psi_{\rm T}={\cal J}\Phi_{0}$, the 
parameters $V_k$ which solve these linear set of equations  will be different 
from zero. 
In order to understand how to use the coefficients $V_k$ to obtain a new set of CI 
coefficients (or $\delta_i$ coefficients in Eq.~\ref{eq:delta}), let us first consider
the case when there is no Jastrow factor and suppose 
that $\Phi_{0}$ is the eigenstate $\Phi_0^{(0)}$ of an {\it incorrect} CI 
Hamiltonian ${\cal H}^{(0)}$:
\begin{eqnarray}
{\cal H}^{(0)}=\sum_k E_k^{(0)}|\Phi_k^{(0)}\rangle\langle\Phi_k^{(0)}|\,,
\label{eq:heff_0}
\end{eqnarray}
where the states $\Phi_i^{(0)}$ are not the CI eigenstates for the given set of 
configuration state functions $C_i$. The coefficients $V_k^{(0)}$ are easily obtained 
from Eq.~\ref{eq:lineq_qmc} as
\begin{eqnarray}
V_k^{(0)}=\langle\Phi_0^{(0)}|{\cal H}|\Phi_k^{(0)}\rangle\,.
\end{eqnarray}
Therefore, the corrections $V_k^{(0)}$ correspond to off-diagonal elements of the 
correct CI Hamiltonian coupling to the lowest state. 
With the use of these coefficients, a new Hamiltonian ${\cal H}^{(1)}$ is constructed 
as
\begin{eqnarray}
{\cal H}^{(n+1)}={\cal H}^{(n)}+\sum_{k>0} V_k^{(n)}(
|\Phi_k^{(n)}\rangle\langle\Phi_0^{(n)}|+h.c.)\,.
\label{eq:update}
\end{eqnarray}
The updated Hamiltonian is diagonalized, yielding a new set of states $\Phi^{(1)}_i$,
and the procedure is repeated until convergence. An iterative approach is necessary
because only the row and the column of the CI Hamiltonian corresponding to $\Phi_0$
are corrected at each step. 
At convergence, one obtains $V_k=\langle\Phi_0|{\cal H}|\Phi_k\rangle=0$ and the
correct CI eigenstate $\Phi_0$.
Therefore, we have devised an iterative scheme to improve
on the starting wave function and perform a CI calculation for $\Phi_0$.

After reintroducing the Jastrow factor, we proceed motivated by the above 
scheme developed in the absence of the Jastrow factor and by the observation 
that, if $\Psi_{\rm T}={\cal J}\Phi_0$ were an eigenstate of the Hamiltonian 
${\cal H}$, $\Phi_{0}$ would be a right eigenstate of the operator 
${\cal J}^{-1}{\cal H}{\cal J}$ with the same eigenvalue. 
Therefore, for Jastrow-Slater wave functions, the EFP method constructs 
iteratively an effective Hamiltonian ${\cal H}_{\rm eff}$ which approximates 
${\cal J}^{-1}{\cal H}{\cal J}$ as far as the action on the trial state 
$\Phi_{0}$ is concerned.
One can then interpret Eq.~\ref{eq:lineq_qmc} as obtained from a least-squares 
fit of the local energy of the effective Hamiltonian ${\cal J}^{-1}{\cal H}{\cal J}$ 
acting on the determinantal part of the trial wave function
\begin{eqnarray}
\chi^2 = \langle 
( \frac{{\cal J}^{-1}{\cal H}{\cal J}\Phi_0}{\Phi_0} - \sum_k V_k O_k )^2  \rangle \,.
\label{eq:chisq_qmc}
\end{eqnarray}
This interpretation allows us to use the corrections $V_k$ to update the approximate effective 
Hamiltonian ${\cal H}_{\rm eff}$ in an iterative scheme as in the absence of the Jastrow factor.

The initial guess for ${\cal H}_{\rm eff}$ is arbitrary but, if we assume that 
an initial CI or MCSCF calculation yields a wave function not too far from the final 
one,  we can construct a reasonable starting Hamiltonian using a complete set of states 
$\Phi_k$ (i.e. as many states as CSF) and the associated energies $E_k$ from such a 
calculation. 
Typically, the QMC trial state $\Phi_{0}$ is truncated according to a threshold 
on the CSF coefficients.   Since the corrections to ${\cal H}_{\rm eff}$ are sampled 
from ${\cal J}\Phi_{0}$, where $\Phi_{0}$ is assumed to be an eigenstate of 
${\cal H}_{\rm eff}$, it is important to keep ${\cal H}_{\rm eff}$ consistent with 
$\Phi_{0}$. To this end, ${\cal H}_{\rm eff}$ is modified to yield the truncated 
$\Phi_{0}$ as an eigenstate by setting to zero the matrix elements between the 
CSF's included in $\Phi_{0}$ and the ones omitted.

As the iterative process proceeds, the effective Hamiltonian will contain more
and more contributions sampled within quantum Monte Carlo and thus
incorporate statistical noise. Consequently, symmetries inherent to the true
Hamiltonian could by broken. This can be avoided by using symmetry adapted
expansion functions instead of simple Slater determinants. If molecular
pointgroups are restricted to Abelian groups, individual determinants of
symmetry adapted orbitals are proper spatial basis functions and only spin symmetry
has to be considered. 

Recently, Umezawa and Tsuneyuki~\cite{transcor} proposed an optimization method
for a single determinant Jastrow-Slater wave function based on the minimization
of the quantity $\int {\rm d}{\bf R}^{3N}\,({\cal H}_{\rm eff} D - E_{\rm guess} D )^2$ 
where ${\cal H}_{\rm eff}={\cal J}^{-1}{\cal H}{\cal J}$. Within this approach, the 
derivatives of the energy with respect to the parameters in the determinant 
are in general not zero upon convergence. 

\subsection{Optimization of Orbitals}

In order to obtain more compact wave functions, it is important to 
optimize the orbitals comprising the
Slater determinants along with the coefficients in front of these
determinants. Within the traditional quantum chemistry framework, this approach
is known as the MCSCF method, which is employed in the generation of reference 
wave functions for both complete-active-space second-order perturbation theory
(CASPT2) and MRCI (multi-reference CI).  
Orbital optimization can be achieved either by using Lagrange multipliers, a 
method similar but much more involved then the SCF method, or using a so-called 
super-CI expansion~\cite{ruedenberg,shepard}. 
Within the latter approach, the original reference space is augmented by all 
possible single excitations with respect to a set of external orbitals, leading 
to the super-CI wave function
\begin{eqnarray}
\Phi_{\rm SCI}=\sum_i \left( c_i C_i + \sum_{kl} \tilde{c}_{kl} C_i^{k\rightarrow l} 
\right)\,.
\end{eqnarray}
The notation $C_i^{k\rightarrow l}$ means that in the reference CSF $C_i$
an electron from the spin-orbital $k$ is promoted to the external spin-orbital
$l$. By virtue of the Brillouin-Levy-Berthier theorem~\cite{BLB}, it can be shown 
that the orbitals of the reference space are optimal with respect to variations in 
the external space if the coefficients in front of appropriate combinations of single
excitations vanish. If convergence has not been reached, the
coefficients $\tilde{c}_{kl}$ can be used to generate improved reference
orbitals. Here, we follow Ruedenberg {\it et al.}~\cite{ruedenberg} in 
generating new orbitals via natural orbitals of the super-CI wave function. 

We reformulate the minimization problem of the orbitals in a Jastrow-Slater 
wave function as a super-CI approach. The resulting procedure 
consists of the following steps:

{\it Step 1}. For a given trial wave function $\Psi_{\rm T}={\cal J}\Phi_{0}$,
all single excitations out of the determinantal component $\Phi_{0}$ are
generated. In this enlarged space of CSF's, the super-CI Hamiltonian is set up 
and modified to yield $\Phi_{0}$ (and not $\Phi_{\rm SCI}$) as an eigenstate. 
Thus, a starting effective super-CI Hamiltonian ${\cal H}^{(0)}_{\rm SCI}$ 
(Eq.~\ref{eq:heff_0}) is obtained with $\Phi_{0}$ as $\Phi^{(0)}_0$.
The quantities appearing in the linear equations (Eq.~\ref{eq:lineq_qmc}) are 
then sampled in a variational Monte Carlo run with wave function $\Psi_{\rm T}$, 
and the corrections $V_k$ are obtained and added to the original Hamiltonian
${\cal H}^{(0)}_{\rm SCI}$. The new effective Hamiltonian 
${\cal H}^{(1)}_{\rm SCI}$ is diagonalized to yield a new state $\Phi^{(1)}_0$.

{\it Step 2}. The natural 
orbitals (i.e.\ the eigenvectors of the single particle density matrix) are computed 
for the state $\Phi^{(1)}_0$. The new reference orbitals are then obtained by
demanding that the natural orbitals of the new reference wave function coincide
with the natural orbitals of the super-CI wave function as explained in 
Appendix~\ref{sec:orbtrafo}.

{\it Step 3}. After the orbital update is performed, the super-CI Hamiltonian 
needs to be recalculated in the basis given by the new orbitals. While this is 
of course possible in the context of usual MCSCF, it is not in the case of the 
EFP since the effective Hamiltonian considered here contains contributions sampled 
within QMC. Instead, we transform the effective Hamiltonian approximately by 
projecting it onto a set of CSF consisting of the new orbitals but having the 
same occupation patterns as the old CSF. 
Subsequent QMC sampling of the matrix elements of the effective Hamiltonian 
as described above will correct this approximation.

Let us denote with $\phi_i$, $C_i$ and $D_i$ 
the old orbitals, CSF's, and determinants, respectively, and with a tilde the 
corresponding new quantities. We project the old set of CSF's onto the new ones 
\begin{eqnarray} 
\tilde{C}_i = \sum_k e_{ik}\; C_{k}\,,
\label{eq:projection}
\end{eqnarray}
where the expansion coefficients are given by $e_{ik}=\langle C_i | \tilde{C}_k
\rangle$.
If the reference space is complete (CASSCF), this relation would be exact.
Otherwise, it leads to a projection of the new effective Hamiltonian onto the new 
set of CSF's $\tilde{C}_i$.

Since the CSF's are linear combinations of determinants (Eq.~\ref{eq:csf})
with the same set of coefficients $d_{ik}$ for the old and new orbitals, the 
expansion coefficients $e_{ik}$ can be evaluated as
\begin{eqnarray}
e_{ik}=\sum_l \; \sum_m \; d_{il}d_{km} \langle D_l | \tilde{D}_k \rangle \,.
\end{eqnarray}
The overlap of two Slater determinants is computed as the determinant of 
the overlap matrix of the spin orbitals occupied in the determinants under 
consideration:
\begin{eqnarray}
\langle D_l | \tilde{D}_k \rangle = 
\det ( \Theta^{\alpha}_{ik} ) \, \det ( \Theta^{\beta}_{ik} )\, ,
\end{eqnarray} 
where $\Theta^{\sigma}_{ik} = \langle \phi^\sigma_i |
\tilde{\phi}^\sigma_k \rangle$, with $\phi^\sigma_i$  the i-th orbital of
spin $\sigma$ in the determinant. 

Finally, the effective Hamiltonian is expressed in the projected basis $\tilde{C}_i$ as
\begin{eqnarray}
\langle \tilde{C}_i | {\cal H}_{\rm eff} | \tilde{C}_j \rangle = \sum_k \sum_l \; e_{ik}
                   e_{jl} \langle C_k | {\cal H}_{\rm eff} | C_l \rangle\,.
\end{eqnarray}

To illustrate the effect of the projection, let us consider a system of two electrons and two
orbitals $\phi_1, \phi_2$ where the reference determinant ('active space') is  $\phi_1^\alpha 
\phi_1^\beta$, and the single excitations into $\phi_2$ ('external space') yield the determinants
$\phi^\alpha_1\phi^\beta_2$ and $\phi^\alpha_2\phi^\beta_1$.  With a new orbital 
$\tilde{\phi}_1=c_1\phi_1+c_2\phi_2$, the (projected) matrix element
of the new reference determinant $\tilde{\phi}^\alpha_1\tilde{\phi}^\beta_1$
is missing the contribution from the determinant $\phi^\alpha_2\phi^\beta_2$
which is proportional to $c_2^2$ and is small if 
$\tilde{\phi}_1$ is close to $\phi_1$.
In general, if the new orbitals do not significantly differ from the old ones
(i.e.\ the occupation numbers of the corresponding natural orbitals are close to 
0, 1, and 2, respectively), the projection approximation is expected to be quite accurate.

\subsection{Multiple states}

Even though, in the above description of the EFP method, we choose the 
trial state $\Psi_{\rm T}$ to be the lowest state ${\cal J}\Phi_0$, this is not a 
requirement. Since the EFP conditions lead to a state for which the energy 
is stationary with respect to parameter variations, arbitrary states could be 
optimized if a close enough starting guess is provided. However, in optimizing a
higher state, one sometimes faces the so-called root flipping problem known 
from CI and MCSCF calculations of excited states: the higher state is lowered
so much in energy that it approaches (and mixes with) an initially lower
state of the same symmetry which is not being optimized. When this happens, the 
procedure either converges to the lower state or, more likely, does not 
converge at all.

Within the MCSCF framework, a well established solution to the root flipping 
problem is the so-called state averaged MCSCF (SA-MCSCF) method~\cite{samcscf}. In
SA-MCSCF, the optimized quantity is the weighted average of the energies 
of the states under consideration:
\begin{eqnarray}
E_{\rm SA}=\sum_{i\in {\rm A}} w_i \frac{\langle \Psi_i | {\cal H} | \Psi_i\rangle}
                              {\langle \Psi_i | \Psi_i\rangle}\,,
\end{eqnarray}
where the weights $w_i$ are fixed and $\sum_i w_i =1$.  
The multi-determinant wave
functions $\Psi_i$ depend on their individual sets of CI-coefficients $c_{ik}$
but on a common set of orbitals. Therefore, if the averaged energy $E_{SA}$ is
stationary with respect to all parameter variations, the individual state
energies $E_i$ are stationary with respect to variations of the
CI-coefficients but {\em not} with respect to variations of the orbitals.
In SA-MCSCF, the wave functions are kept orthogonal and a generalized variational 
theorem applies~\cite{samcscf}.
 
The concept of state averaging is introduced into the EFP framework by 
considering several states $\Psi_i, i \in A$, and associated weights 
$w_i$ ($\sum_i w_i=1$) instead of a single trial state $\Psi_{\rm T}$. 
In the EFP procedure, one needs to modify how  the corrections to the
effective Hamiltonian are obtained and how the natural orbitals are computed. 

For multiple states, we start from a least squares
problem like in Eq.~\ref{eq:chisq}, but with the single trial function replaced by the
ensemble $\Psi_i={\cal J}\Phi_i, i \in A$, with the same Jastrow factor for all states:
\begin{eqnarray}
\chi^2 
& = & \sum_{i\in A} w_i \langle ( E_L^{(i)} - \sum_k V_{ki} O_{ki} )^2  \rangle_i \;,
\label{eq:chisq_av}
\end{eqnarray}
where the local energy and the functions $O_{ki}$ are defined as before but refer now to the
different trial states,
\begin{eqnarray}    
E_L^{(i)}=\frac{{\cal H} \Psi_i}{\Psi_i} \quad\text{and}\quad O_{ki}=\frac{\Phi_k}{\Phi_i} 
\end{eqnarray}
and $\langle\cdot\rangle_i$ denotes the average with respect to $|\Psi_i|^2$. 
Since the optimization of orbitals is expressed as optimization of CI coefficients in 
the external space, all parameters become state-specific and the minimization of $\chi^2$ 
leads to a different set of linear equations for each state.
The corrections $V_{ki}$ from these equations are then combined with the weights $w_i$ to 
yield a new single approximate effective Hamiltonian.   

After this averaged effective Hamiltonian has been  diagonalized, a set of natural orbitals common to all states is
obtained from the averaged single particle density matrix
\begin{eqnarray}
\rho_{\rm SA}= \sum_{i \in {\rm A}} w_i \rho_i \ .
\end{eqnarray}
As a result of this last step, the corrections $V_{ki}$ corresponding to 
external excitations do not vanish upon convergence and, as in the SA-MCSCF, only 
the averaged energy will become stationary with respect to orbital variations. 

Since our procedure imposes orthogonality only among the determinantal parts $\Phi_i$, 
the full wave functions $\Psi_i$ will in general not be orthogonal due to the presence 
of the Jastrow factor. Therefore, the generalized variational theorem providing lower 
bounds for the excited state energies can be violated.  
  
\subsection{\label{sec:sr}The stochastic reconfiguration approach to wave function optimization}

We briefly review the so-called stochastic reconfiguration (SR) approach~\cite{sr1,sr2}
of Sorella {\it et al.} and discuss similarities and differences with the
EFP method.

In the SR method, an improved state is obtained by applying the operator
$\Lambda-{\cal H}$ to the current trial state
\begin{eqnarray}
\Psi^{(n+1)} = (\Lambda-{\cal H}) \Psi^{(n)}
\label{eq:lanczositer}
\end{eqnarray}
where $\Lambda$ is an energy shift which controls the rate of convergence.

If there is a parameterized {\em ansatz} for $\Psi^{(n)}$ with parameters
$\alpha_k, k=1 \dots p$, a linear variation with respect to all parameters can
be written as
\begin{eqnarray}
\Psi^{(n+1)} = \sum_{k=0}^p \delta_k O_k \Psi^{(n)} \ ,
\label{eq:ansatz}
\end{eqnarray}
where the quantities $O_k$ with $k>0$ are the logarithmic derivatives of $\Psi^{(n)}$
with respect to the parameter $\alpha_k$
\begin{eqnarray}
O_k = \frac{\partial}{\partial \alpha_k} \ln \Psi^{(n)} \ ,
\end{eqnarray}
and $\delta_k$ denotes a change in the parameter $\alpha_k$.  The operator
$O_0$ is the identity and the associated parameter $\alpha_0$ corresponds to
an overall scaling of the wave function.

The central idea of the SR approach is to apply $\Lambda-{\cal H}$ to the
current trial wave function and project the result onto the space defined by
the parameterization. This leads to the conditions
\begin{eqnarray}
\langle \Psi^{(n)} | O_k (\Lambda-{\cal H}) | \Psi^{(n)} \rangle = \langle
\Psi^{(n)} | O_k | \Psi^{(n+1)} \rangle.
\end{eqnarray}
After inserting Eq.~\ref{eq:ansatz} for $\Psi^{(n+1)}$ and replacing
$\Psi^{(n)}$ with $\Psi_{\rm T}$, the following equations are obtained for
$k=0 \dots m$:
\begin{eqnarray}
\langle O_k (\Lambda - E_L) \rangle = 
\sum_{l=0} \delta_l \langle  O_k O_l \rangle \,.
\end{eqnarray}
The $k=0$ equation yields the scaling $\delta_0$ in terms of $\Lambda$:
\begin{eqnarray}
\delta_0 = \Lambda - \bar{E} - \sum_{l>0} \delta_l \langle O_l \rangle.
\end{eqnarray}
After substitution of $\Lambda$, the equations with $k>0$ become
\begin{eqnarray}
\bar{E} \langle O_k \rangle - \langle O_k E_L \rangle = \sum_{l>0} \delta_l (\langle
O_k O_l \rangle - \langle O_k \rangle \langle O_l \rangle ) \,,
\end{eqnarray}
which can be rewritten as
\begin{eqnarray}
- \langle \Delta O_k \Delta E\rangle = \sum_{l>0} \delta_l \langle \Delta O_l
\Delta O_k\rangle \ .
\label{eq:lanczos}
\end{eqnarray}
 These equations coincide, apart from the minus sign, with the working equations of 
the EFP method (Eq.~\ref{eq:lineq_qmc}). However, they describe different quantities: changes in the wave function
($\delta_l$) in the SR method, and changes to the Hamiltonian ($V_l$) in the EFP method.

To investigate the connection between the EFP and the SR method, let us suppose that 
the new eigenstates of the EFP effective Hamiltonian are not obtained through 
diagonalization but approximately to first order in the perturbation given by 
the corrections $V_l$. Then, the change of the trial state with respect to the
$k$-th eigenstate $\Phi_k$ of the old effective Hamiltonian is given by
\begin{eqnarray}
\Psi_{\rm T} = {\cal J} \Phi_{0} \rightarrow {\cal J} \left(\Phi_{0} - 
\frac{V_k}{E_k-E_{0}}\Phi_k\right)\,,
\end{eqnarray}
while, in the SR method, the same quantity is obtained as 
\begin{eqnarray}
\Psi_{\rm T} = {\cal J} \Phi_{0} \rightarrow {\cal J} \left(\Phi_{0} + \frac{\delta_k}{\delta_0}
\Phi_k\right)\,,
\end{eqnarray}
where the parameter $\delta_k$ corresponds to a variation with respect to the eigenstate $\Phi_k$.
Therefore, the main difference between the two methods seems to be a {\em parameter
specific scaling} in the EFP as opposed to a global one in the SR method. We found this 
difference to affect the convergence rate of the methods sometimes considerably, as will 
be shown in Sec.~\ref{sec:results}. 
 
As pointed out by Sorella {\it et al.}, there is a relationship between the choice of the
energy shift parameter $\Lambda$ and the amount of sampling to determine the
corrections $\delta_k$ in each iteration. The more accurately these are
sampled, the smaller one can make $\Lambda$, achieving faster convergence.
However, as shown in Appendix \ref{sec:srconverg}, a simple consideration of the 
convergence behavior suggests that there exists a critical $\Lambda_c$ below which 
convergence can not be achieved, a finding which is in agreement with our numerical results.

\section{\label{sec:comp}Computational details}

The vertical excitations of ethene are computed using the experimental 
ground state geometry which is of D$_{2h}$ symmetry ($R_{\rm CH}=1.086 \AA$, 
$R_{\rm CC}=1.339 \AA$ and $\angle{\rm HCH}=117.6^\circ$)~\cite{lischka}. 
The molecule is placed in the $yz$- plane with the molecular axis along the $z$ direction.

The carbon 1$s$ electrons are 
replaced by a norm-conserving $s$-non-local pseudopotential generated in an all-electron 
Hartree-Fock calculation for the carbon atom. The potential of the hydrogen atom is 
softened by removing the Coulombic divergence.

The Gaussian basis sets are optimized for our soft pseudopotentials and augmented 
with polarization and diffuse functions. The calculations are performed with two
different basis sets. Basis (A) is a contracted (12$s$12$p$2$d$)/[5$s$5$p$2$d$]
basis with the most diffuse exponent being 0.02. Basis (B) consists of basis (A)
augmented with two more diffuse $s$ and $p$ functions with exponents 0.005 and 
0.002.

The Hartree-Fock and CASSCF calculations are performed with the program 
{\tt GAMESS(US)}~\cite{gamess}. In all SA-MCSCF calculations, equal weights
are employed for the two states. 
The program package {\tt CHAMP}~\cite{champ} is used for the quantum Monte Carlo 
calculations.  Different Jastrow factors are used to describe the correlation with a 
hydrogen and a carbon atom, and their parameters are optimized 
within QMC using the variance minimization method~\cite{varmin}.
We employ both a `2-body' Jastrow factor consisting of electron-electron and 
electron-nucleus terms, and a `3-body' Jastrow factor where additionally 
electron-electron-nucleus terms are included~\cite{jastrow}.  
An imaginary time step of 0.075 H$^{-1}$ is used in the DMC calculations.

Singular value decomposition with a threshold of $0.0001$ is used for inverting the
matrix $\left<\Delta O_l \Delta O_k \right>$ in  Eq.~\ref{eq:lineq_qmc}. We always find a large gap
in the spectrum of this matrix so that the results are not sensitive to the precise value of the
threshold. 
 
\section{\label{sec:results}Results and Discussion} 

All traditional quantum chemistry as well as QMC techniques rely on the fact that 
their reference or trial wave function captures the essential nature of the state
under consideration, whose description is then refined by including
dynamic correlation. 
This precisely fails in the case of the valence $1{}^1\mbox{B}_{1u}$ state of ethene
since explicit inclusion of dynamic correlation already in the reference seems 
necessary to avoid mixing with Rydberg states. 
Moreover, the problem worsens with increasing single-particle basis set,
especially with the addition of more diffuse functions,
since a simple reference wave function yields a state of even more Rydberg 
nature. 

An additional complication with this state is 
posed by the occupied $\sigma$ orbitals which respond to the $\pi\rightarrow\pi^*$ 
excitation and cannot be treated as frozen.
Although ethene seems to be a very special case, similar problems
are expected to be present in many photoactive molecules, possibly in a 
milder but also less clear-cut way. 
For all these reasons, the $1{}^1\mbox{B}_{1u}$ state of ethene represents 
a very stringent test for methods which aspire to provide highly accurate excitation 
energies, and is therefore chosen here as an illustrative example to demonstrate the 
effectiveness of our EFP optimization method.

Although the $1{}^1\mbox{B}_{1u}$ state of ethene is an ideal test case for 
sophisticated correlation methods, a down-side must be mentioned: 
it is very difficult to extract reliable estimates of this vertical 
excitation from gas phase experiments since ethene starts immediately to twist upon 
photoexcitation.  Current interpretations of the measured data seem to indicate 
7.7~eV as a lower bound to the $1{}^1\mbox{B}_{1u}$ vertical excitation energy
of ethene~\cite{mcdiarmid,davidson2}.
Theoretical excitation energies obtained with contemporary 
quantum chemistry methods vary between 7.69~eV from 
MRCI~\cite{lischka} and 8.4~eV from CASPT2~\cite{roos}.

\begin{table*}[htb]
\caption{Total VMC and DMC energies in Hartree and spatial extent of the wave function $\langle X^2 \rangle$ 
in $a_0^2$ for the states $1{}^1\mbox{A}_g$,  $1{}^1\mbox{B}_{1u}$, $2{}^1\mbox{A}_g$, 
$2{}^1\mbox{B}_{1u}$ and $3{}^1\mbox{B}_{1u}$ of ethene. Different combinations of basis sets
and active spaces are used. A 2-body Jastrow factor is employed unless indicated as '3body-J'. 
The total number of occupied orbitals in the reference is listed for each wave function type, 
together with the number of optimized orbitals and variational parameters.
The DMC excitation energies in eV are computed with respect to the ground state energy obtained 
with the same basis and Jastrow factor as in the excited state. We do not report the ground state 
DMC energies for basis (B) for reoptimized orbitals or a 3-body Jastrow factor since
the same behavior is observed as for basis (A).
The numbers in parentheses are the statistical errors.
See text for a detailed explanation.
\label{table:ethylene1}}
\begin{ruledtabular}
\begin{tabular}{lllllllllll}
state & basis & wave function & occupied orb. & optimized orb. & parameters & $E_{\text{vmc}}$ &  
$\langle X^2 \rangle $ &   $E_{\text{dmc}}$ & $ \Delta E$ (eV) \\\hline     
$1{}^1\mbox{A}_g$       & A &   HF       & 6  & 0    &  -   &  -13.6744(5) & 12 & -13.7194(4) &    -    \\
                        &   &            &    & 6    & 101  &  -13.6797(5) & 12 & -13.7204(4) &    -    \\
                        &   & 3body-J    &    & 6    & 101  &  -13.6935(5) & 11 & -13.7203(4) &    -    \\
\cline{2-10}
                        & B &   HF       & 6  & 0    &  -   &  -13.6737(5) & 12 & -13.7191(4) &    -    \\\hline
$1{}^1\mbox{B}_{1u}$    & A & CAS 2-2    & 7  & 0    &   -  &  -13.3502(5) & 37 & -13.4095(4) & 8.45(2) \\  
                        &   &            & 7${}^a$  & 0    &   -  &  -13.3321(5) & 12 & -13.4116(4) & 8.40(2) \\ 
                        &   &            &    & 2    & 17   &  -13.3694(5) & 20 & -13.4245(4) & 8.05(2) \\
                        &   &            &    & 7    & 201  &  -13.3738(5) & 20 & -13.4257(4) & 8.02(2) \\
                        &   & 3body-J    &    & 7    & 201  &  -13.4011(5) & 19 & -13.4292(4) & 7.92(2) \\
\cline{3-10}
                        &   & CAS 6-6    & 9  & 0    &  -   &  -13.3546(5) & 32 & -13.4134(4) & 8.35(2) \\ 
                        &   &            &    & 9    & 858  &  -13.3803(5) & 18 & -13.4280(4) & 7.95(2) \\
                        &   & 3body-J    &    & 9    & 858  &  -13.4008(5) & 19 & -13.4285(5) & 7.94(2) \\
\cline{2-10}
                        & B & CAS 2-2    & 7  & 0    &  -   &  -13.3517(5) & 52 & -13.4091(4) & 8.47(2) \\ 
                        &   &            &    & 7    & 233  &  -13.3739(5) & 22 & -13.4245(4) & 8.05(2) \\ 
                        &   & 3body-J    &    & 7    & 233  &  -13.4004(5) & 24 & -13.4289(4) & 7.93(2) \\\hline
$2{}^1\mbox{A}_{g}$     & A & CAS 2-2    & 7  & 0    & -    &  -13.3687(5) & 52 & -13.4279(4) & 7.96(2) \\ 
                        &   &            &    & 7    & 293  &  -13.3708(5) & 46 & -13.4129(4) & 8.36(2) \\
                        &   & 3body-J    &    & 7    & 293  &  -13.3868(4) & 49 & -13.4129(4) & 8.36(2) \\\hline
$2{}^1\mbox{B}_{1u}$    & A & CAS 2-3    & 8  & 0    & -    &  -13.3058(5) & 42 & -13.3877(4) & 9.05(2) \\
                        &   &            &    & 8    & 401  &  -13.3136(5) & 49 & -13.3699(4) & 9.53(2) \\
                        &   & 3body-J    &    & 8    & 401  &  -13.3439(4) & 52 & -13.3758(4) & 9.37(2) \\
\cline{3-10}
$1{}^1\mbox{B}_{1u}$ (SA)${}^b$  &   & CAS 2-3 & 8  & 8    & 401  &  -13.3718(5) & 22 & -13.4233(4) & 8.08(2) \\
                                 &   & 3body-J &    & 8    & 401  &  -13.3994(5) & 20 & -13.4279(4) & 7.95(2) \\\hline
$1{}^3\mbox{B}_{1u}$ & A & CAS 2-2    & 7  & 7    & 293  &  -13.5118(5) & 12 & -13.5553(4) & 4.49(2) \\
\end{tabular}
\end{ruledtabular}

$a$ Two-determinant wave function with triplet orbitals.\\
$b$ This state is obtained in a state-average (SA) calculation with the corresponding $2{}^1\mbox{B}_{1u}$
state.
\end{table*}

\begin{table}[htb]
\caption{Comparison of vertical excitation energies of ethene from fixed-node DMC with experiment and 
other theoretical studies. All excitation energies are in eV. 
\label{table:ethylene2}}
\begin{ruledtabular}
\begin{tabular}{llllll}
State         &    DMC                      &  Exp.                  &  MRCI                                       \\\hline
$1{}^1B_{1u}$ &    7.93(2)                  &  $>$7.7${}^a$          &  7.69${}^b$, 7.8 ... 7.9${}^c$, 7.96${}^d$  \\
$1{}^3B_{1u}$ &    4.49(2)  4.50(3)${}^e$   &  4.36${}^d$,           &  4.49${}^d$                                 \\
$2{}^1A_g$    &    8.36(2)                  &  8.29${}^d$            &  8.21${}^d$                                 \\ 
$2{}^1B_{1u}$ &    9.37(2)                  &  9.33${}^d$            &  9.17${}^d$                                 \\
\end{tabular}
\end{ruledtabular}

$a$  Ref. \onlinecite{mcdiarmid},
$b$  Ref. \onlinecite{lischka},
$c$  Ref. \onlinecite{buenker},
$d$  Ref. \onlinecite{wiberg},
$e$  Ref. \onlinecite{lester}

\end{table}

In Table~\ref{table:ethylene1}, we present the VMC and fixed-node DMC energies 
for the ground state $1{}^1\mbox{A}_g$ and the $1{}^1\mbox{B}_{1u}$ state of ethene.
We also list the VMC 
expectation value of the spread of the wave functions in the direction perpendicular to the 
molecular plane, $\langle X^2 \rangle=\langle \Psi_{\rm T}| \sum_i x_i^2 |  \Psi_{\rm T}\rangle$,
since this quantity has proven useful to differentiate between the valence and Rydberg 
nature of a state. Additionally, we show results for the higher singlet 
states  $2{}^1\mbox{A}_g$ and $2{}^1\mbox{B}_{1u}$ and the $1{}^3\mbox{B}_{1u}$ triplet state.  

\begin{figure}[htb]
\includegraphics[width=8.5cm]{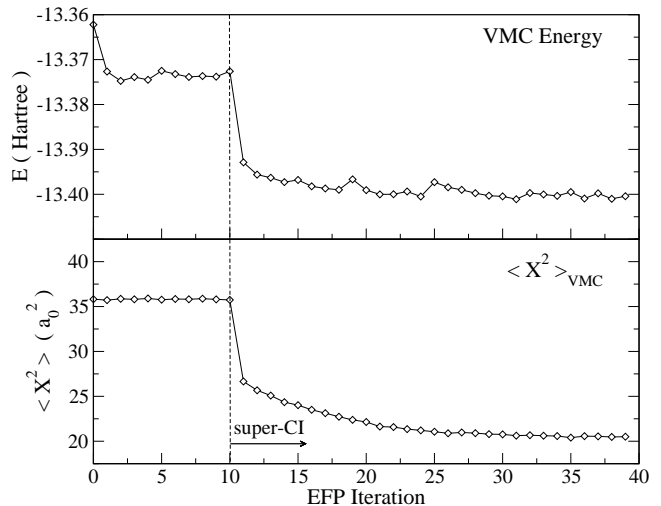}
\caption{\label{fig:iter1} Example of the convergence of the VMC energy and the spatial
extent $\langle X^2 \rangle$ of the wave function in the EFP optimization of a CAS 6-6
wave function with basis (A). Initially, only the CI coefficients are 
being optimized. After the first 10 iterations, a threshold of 0.01 is applied to the 
coefficients of the CSF's and only single excitations on top of this truncated CI wave
functions are considered to optimize all the occupied orbitals. 
The statistical error is smaller than the size of the symbols.
See the text for more details.}
\end{figure}

For the $1{}^1\mbox{B}_{1u}$ state, we investigate how the QMC excitation energy varies
when going from basis set (A) to the more diffuse basis set (B), and whether our EFP 
optimization method is able in both cases to correct the Rydberg character of the starting MCSCF 
reference used in the Jastrow-Slater wave function. 
Since the $1{}^1\mbox{B}_{1u}$ state is expected to have a pronounced HOMO-LUMO character, we first 
employ a simple two-determinant wave function, corresponding to a 2 electron
in 2 orbitals CASSCF wave function (denoted by 'CAS 2-2').
To improve on possible deficiencies of this description, we also consider a 6 electron 
in 6 orbitals CASSCF wave function (denoted by 'CAS 6-6') where the active space consists of 
the orbitals $1b_{2g}$, $1b_{3u}$, $2a_g,1b_{3g}$, $2b_{2u}$, and $2b_{1u}$. 
Since optimizing  both CI coefficients and orbitals for this active space would result
in more than $10^5$ CSF's which is not feasible with our present implementation, we first 
optimize the CI coefficients with respect to a fixed set of orbitals. Upon convergence, we 
apply a threshold of 0.01 to the coefficients of the CSF's and augment the resulting truncated 
CI expansion with single excitations in order to relax the orbitals.  
In this last step, single excitations from 'core' orbitals, i.e.\ occupied orbitals that 
were not included in the CAS, can be considered as well. 
The number of virtual orbitals for single excitations is dictated by 
technical limitations: while, for the CAS 2-2  wave function, all virtual orbitals
can be used, in the case of the truncated CAS 6-6 wave function, we include 
the lowest 50 virtual orbitals.
An illustration of the two-step procedure is given in Fig.~\ref{fig:iter1}: after the initial 
optimization of the CI coefficients, the orbitals in the reference comprised by the relevant
CSF's are relaxed. 
This two-step approach is justified by the very different role of the reference wave function 
in QMC compared to traditional quantum chemistry methods: a smaller number of determinants is 
needed in a Jastrow-Slater wave function since the reference does not define the 
available excitation space for the description of dynamical correlation as it is the 
case for a method like MRCI.
Finally, we investigate the effect on the excitation energy as an increasing number 
of occupied orbitals are reoptimized, and  the impact of using a 3-body versus 
a 2-body Jastrow factor.

The results in Table~\ref{table:ethylene1} for the $1{}^1\mbox{B}_{1u}$ state can be 
summarized as follows.
The initial MCSCF Jastrow-Slater wave functions have a substantial Rydberg contribution 
which, as expected, increases when the basis set becomes more diffuse. The spread 
$\langle X^2 \rangle$ in the direction perpendicular to the molecular plane increases 
from 37 $a_0^2$ with basis (A) to 52 $a_0^2$ with basis (B) for the CAS 2-2
wave function.
The resulting fixed-node DMC excitation energies are about 8.5~eV, thus much higher 
than what was found in other benchmark {\em ab initio} calculations. They roughly
agree with CASPT2 results which are believed to suffer from the same deficiency.    

Optimizing the orbitals in the presence of the Jastrow factor, i.e.\ including the
feedback of dynamic correlation on the reference, reduces the spread of the wave function 
to 18-22 $a_0^2$, slightly larger than the ground state value.
The DMC excitation energies are lowered by as much as 0.5-0.6~eV. 
After optimization, the results obtained with basis A and basis B are indistinguishable. 

For CAS 2-2 wave functions, the most substantial reduction in the VMC and DMC energies and in 
$\langle X^2 \rangle$ is obtained when the active orbitals are optimized in the presence of 
the Jastrow factor.
If $\sigma$ orbitals are included in the optimization, further smaller improvements are gained 
at the variational level while the DMC energies remain rather insensitive. Using a CAS 6-6 wave 
function instead of the two-determinant wave function yields lower VMC energies but, for fixed-node 
DMC, the effect of the multi-configuration trial wave function is not very large. In 
fact, after the optimization, there remain only 10 determinants with CI coefficients above our 
threshold of 0.01 while, in the initial CAS 6-6 wave function from MCSCF,
36 determinants meet the same criterion. 
When a 3-body Jastrow factor is employed, the difference between a CAS 6-6 and a CAS 2-2 
energy becomes negligible also at the VMC level.
In general, the use of a 3-body instead of a 2-body Jastrow factor always gives lower VMC 
energies and lower excited-state DMC energies~\cite{non_loc_pseudo} while the DMC energies 
of the ground state are rather unaffected. The resulting improvement on the DMC excitation 
energy is visible only when a CAS 2-2 Jastrow-Slater wave function is used to localize
the pseudopotential~\cite{non_loc_pseudo}.

The reduction of the spatial extent of the trial wave function after reoptimization 
to values around 20 $a_0{}^2$ is in qualitative agreement with the findings 
of Buenker {\em et al.}~\cite{buenker} and Lischka {\em et al.}~\cite{lischka}
although we observe a considerably larger spread of the values depending on the 
details of the optimization. As Fig.~\ref{fig:iter1} illustrates, the convergence of 
$\langle X^2 \rangle$ with the number of steps in the EFP optimization is much slower 
than the convergence of the energy and harder to establish. Interestingly, the EFP
optimization appears to be sensitive to quantities other than the energy: even though the
energy is practically converged, additional optimization steps yield a systematic
lowering of $\langle X^2 \rangle$.

Our final DMC results are summarized and compared with experiment and other theoretical studies in 
Table~\ref{table:ethylene2}.
The best estimate for the
excitation energy of the  $1{}^1\mbox{B}_{1u}$ state is $7.92 \pm 0.02$~eV, in good
agreement with most other {\em ab initio} calculations~\cite{buenker,wiberg}. 
However, it is higher by about 0.23~eV than the value of 7.69~eV recently obtained by Lischka
{\em et al.} using a sophisticated MRCI approach with up to 12 electrons in 12
orbitals reference spaces. It should be noted that such MRCI calculations suffer from a 
size-consistency problem which can be approximately corrected using a term proposed by Davidson
(see e.g. Ref.~\onlinecite{lischka}). For the present example, the Davidson correction 
amounts to about 0.1~eV and Buenker and Krebs questioned its trustworthiness~\cite{buenker}. 

The MRCI results of Buenker and Krebs were partly obtained using MCSCF triplet orbitals which are
as compact as the ground state orbitals. Following this established strategy, recent DMC calculations 
also employed triplet orbitals in trial wave functions for singlet states~\cite{porphyrin}.
However, when we use this recipe for ethene, the corresponding fixed-node 
DMC value is $8.4 \pm 0.02$ eV, which is comparable with the value obtained from singlet MCSCF orbitals.

In the case of the $2{}^1\mbox{A}_{g}$ and $2{}^1B_{1u}$ states, we employ 
our state-averaged EFP optimization approach because these states are not the lowest in 
their irreducible representation and, therefore, have to be optimized in the presence of the 
$1{}^1\mbox{A}_{g}$ and $1{}^1B_{1u}$ states, respectively. In DMC, we rely instead
on the fixed-node approximation to prevent a collapse to the lower states (see 
Appendix~\ref{sec:dmc}). The EFP optimization of multiple states is found to be quite stable:
both VMC and DMC energies of the 
$1{}^1B_{1u}$ state taken from the state-averaged optimization (denoted by 'SA' 
in Table~\ref{table:ethylene1}) are in agreement with the corresponding values 
obtained by optimizing the lower state alone.  
Moreover, upon optimization, the DMC excited state energies are substantially higher than
those obtained using MCSCF orbitals which give excitation energies lower
than the experimental values by 0.33~eV for the $2{}^1\mbox{A}_{g}$ and 0.48~eV for the
$2{}^1B_{1u}$ state. This can possibly be understood from the fact
that valence-Rydberg mixing in the initial MCSCF wave function raises the lower 
state and lowers the higher one.  
As in the case of the $1{}^1B_{1u}$ state, the best excitation energy for the $2{}^1B_{1u}$ state
is obtained when a 3-body Jastrow factor is employed, while the $2{}^1\mbox{A}_{g}$ excitation
is insensitive to the change from a 2-body to a 3-body Jastrow factor.
Our final DMC excitation energies of $8.36 \pm 0.02$~eV for $2{}^1\mbox{A}_{g}$ and 
$9.37 \pm 0.02$~eV for $2{}^1B_{1u}$ are only slightly higher than the corresponding 
experimental values~\cite{wiberg} of $8.29$~eV and $9.33$~eV, respectively.
 
For the  $1{}^3B_{1u}$ state we obtain an excitation energy of $4.49 \pm 0.02$~eV 
which agrees with the recent DMC result~\cite{lester} of El 
Akramine {\em et al.},  the experimental value quoted in 
Ref.~\cite{wiberg} and the MRCI calculation of Krebs and Buenker~\cite{buenker}.
The higher  MRCI value of $4.61$eV reported by Gemein and 
Peyerimhoff~\cite{gemein} seems to result from using a somewhat different geometry.
 
\begin{figure}
\includegraphics[width=8.5cm]{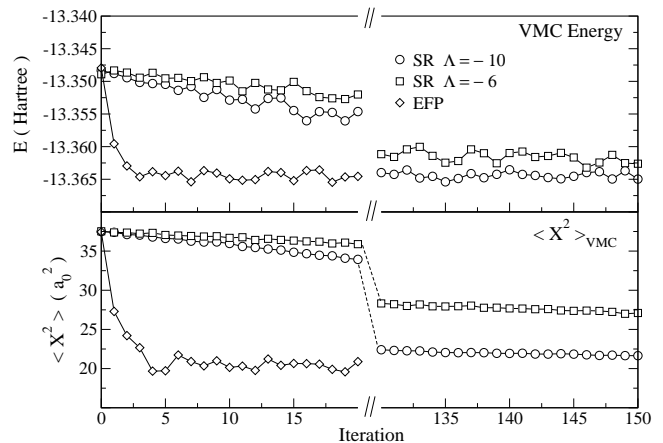}
\caption{\label{fig:iter2} Convergence of the VMC energy and the spatial extent 
$\langle X^2 \rangle$ of the ethene $1{}^1B_{1u}$ state with the iteration number
in the optimization.
The EFP and the SR method with two different values of the parameter $\Lambda$
are used to optimize the two active orbitals in a two-determinant wave function
with basis (A).
For values of $\Lambda$ lower than -10 a.u., the SR optimization is unstable. 
The statistical error is smaller than the size of the symbols.
See text for more details. }
\end{figure}

Finally, we examine the differences between our EFP approach and the stochastic
reconfiguration method by Sorella {\it et al.}. As explained in Sec.~\ref{sec:sr}, 
the SR method applied to the optimization of the determinantal part of the trial wave 
function can be considered as a simpler variant of the EFP approach that should lead 
to the same solution upon convergence but may have different convergence properties. 
In Fig.~\ref{fig:iter2}, the different performance of the two approaches  is illustrated 
with the optimization of the two active orbitals in a two-determinant basis-(A) wave 
function of the $1{}^1\mbox{B}_{1u}$ state of ethene. 
Both calculations are performed using the same amount of Monte Carlo sampling 
per iteration. While the EFP method is parameter-free, the convergence of the 
SR approach depends on the choice of the parameter $\Lambda$. We 
find the optimal value of $\Lambda$ leading to the fastest convergence to be 
about $-10$ a.u., a value which approximately agrees with our simple estimate of the
critical value $\Lambda_c$ given in Appendix~\ref{sec:srconverg}.
Values of $\Lambda$ higher than $\Lambda_c$ yield slower convergence as can be 
seen in Fig.~\ref{fig:iter2}. On the other hand,
decreasing $\Lambda$ below $\Lambda_c$ results in the divergence of the energy
since the contributions of some excited states in the wave function start to be 
amplified instead of damped. 
While the EFP optimization is essentially converged after about 10 iterations, the SR 
approach even at the optimal value of $\Lambda=-10$ a.u.\ takes more than 100
iterations to converge.
As explained in Appendix~\ref{sec:srconverg}, the critical 
value $\Lambda_c$ is related to the energy spread inside the variational space which, 
in the case of orbital variations, is the space spanned by the single excitations 
from the reference configurations into the set of admitted virtual orbitals. 
For basis (A) and a two-determinant reference, if we use all virtual orbitals of 
proper symmetry, this spread amounts to about $6$ a.u. 
If we reduce the energy spread by excluding high-lying virtual orbitals which are 
not expected to contribute significantly to the optimization, $\Lambda$ can be further 
decreased, yielding a faster convergence. 
However, the convergence of the SR approach is expected to become slower as
the size of the system increases since the energy spread of the variational 
space will increase as well.

\section{\label{sec:conclusion}Conclusions}

We extended the energy fluctuation potential (EFP) method to simultaneously
optimize the orbitals and the CI coefficients in Jastrow-Slater wave 
functions via a super-CI approach, and to treat state-averaging for the 
optimization of multiple states of the same symmetry. With these additions, the 
method becomes a useful and effective tool to optimize trial wave functions for 
both ground and excited states, which can then be used in fixed-node diffusion 
Monte Carlo. 

As illustrative examples, we considered several vertical excitations of ethene, 
in particular the difficult valence $1{}^1B_{1u}$ state. For this state, the EFP 
approach leads to strongly improved trial wave functions upon the starting 
Rydberg-like MCSCF reference, as can especially be seen from the reduced spatial
spread of the optimized wave function. All results seem in reasonable agreement with
experiment and other quantum chemical methods. In contrast to techniques like MRCI and CASPT2,
it is not necessary to use multi-configuration trial wave functions once the orbitals
are optimized in the presence of the Jastrow factor. Also, using triplet orbitals 
for singlet states does not appear to be a reliable recipe for fixed-node DMC.

Our EFP optimization of multiple states was employed for the two Rydberg-like states 
$2{}^1\mbox{A}_{g}$ and $2{}^1\mbox{B}_{1u}$ of ethene, which are not the lowest ones in 
their irreducible representations. The optimization procedure is stable and yields
substantially higher DMC excited state energies than the ones obtained with MCSCF
orbitals, bringing the corresponding excitation energies in better agreement with 
the experimental values.

In addition, we also compared the EFP method to the stochastic reconfiguration 
method of Sorella {\em et al.}~\cite{sr2}. These two methods stem from quite different 
theoretical backgrounds but, for practical purposes, the EFP approach can be regarded 
as a more sophisticated variant of the SR scheme. In general, the EFP method shows much faster 
convergence than the SR approach.
On the other hand, 
the SR technique can be applied to the optimization of arbitrarily parameterized  wave 
functions while the EFP method is always based on a suitable factorization of the wave function.

\acknowledgements
This work is part of the research programme of the Stichting voor Fundamenteel Onderzoek der 
Materie (FOM), which is financially supported by the Nederlandse Organisatie voor Wetenschappelijk
Onderzoek (NWO).

\appendix
\section{\label{sec:orbtrafo}Orbital transformation}
Let $U$ be the orthogonal transformation between the current orbitals $\phi$
and the corresponding natural orbitals $\phi^{\text{nat ref}}$ of the reference wave function $\Phi$:
\begin{eqnarray}
\phi^{\text{nat ref}}_i & = & \sum_{j=1}^M U_{ij} \phi_j 
\end{eqnarray}
After updating and rediagonalizing the super-CI Hamiltonian (i.e.\ reference plus 
single excitations), the natural orbitals of this wave function are given by the 
transformation $V$: 
\begin{eqnarray}
\tilde{\phi}^{\text{nat ext}}_i & = &
\sum_{j=1}^N V_{ij} \phi_j
\end{eqnarray}
Here, $M$ and $N$ ($N \gg M$) denote the dimensions of the reference space and 
the super-CI space, respectively.
The new reference orbitals $\tilde{\phi}$ are now obtained as
\begin{eqnarray}
\tilde{\phi}_i=\sum_{j=1}^M \sum_{k=1}^N U_{ji} V_{jk} \phi_k \quad i=1\dots
M\ ,
\end{eqnarray}
where the orthogonality of $U$ ($U^{-1}=U^T$) was used.  
Upon convergence, this ensures that the natural orbitals of the reference wave function 
coincide with the corresponding subset of the natural orbitals of the super-CI expansion.

As the new virtual orbitals ($\tilde{\phi}_i, i=M+1\dots N$) one can use the orbitals 
$\phi^{\text{nat ext}}_i , i=M+1\dots N$ since they are orthogonal to the occupied ones. 
Alternatively, one can obtain a new set of virtual orbitals by explicit orthogonalization 
of the orbitals of the previous iteration to the new occupied orbitals.

It should be noted that in the case of a complete reference space (CAS) the
natural orbitals of the super-CI expansion could be used instead of the orbitals 
from the transformation above, since the CAS wave function is invariant with respect 
to transformations among the active orbitals. However, subsequent truncation of
the wave function (omission of determinants with coefficients below some
threshold) could lead to slightly different results depending on the orbital
transformation employed.

\section{\label{sec:srconverg}Critical scaling parameter of the SR method}

Let us suppose that the $M$ eigenstates of some Hamiltonian ${\cal H}$ are known
\begin{eqnarray}
{\cal H} \Psi_i = E_i \Psi_i, \;\;\; i=0, \dots, M \ .
\end{eqnarray} 
An approximation $\Psi$ to the state $\Psi_0$ can be written  as 
\begin{eqnarray}
\Psi= \Psi_0 + \sum_{i>0} c_i \Psi_i \ .
\end{eqnarray} 
The variations with respect to the states $i>0$ are introduced as
\begin{eqnarray}
\Psi^\prime  = \Psi + \sum_{i>0} \delta_i \Psi_i 
 =  \Psi_0 + \sum_{i>0} c_i \alpha_i \Psi_i \ .
\end{eqnarray} 
Acting with the operator $\Lambda-{\cal H}$ on $\Psi$ yields
\begin{eqnarray}
\tilde{\Psi} & = & (\Lambda-{\cal H}) \Psi \nonumber \\ 
& = & (\Lambda-E_0)\Psi_0 + \sum_{i>0} c_i (\Lambda - E_i) \Psi_i \ .
\end{eqnarray} 
Therefore, after dividing by $\Lambda-E_0$ and equating $\Psi^\prime$ 
with $\tilde{\Psi}$, we obtain that the corrections $\alpha_i$ are given by
\begin{eqnarray}
\alpha_i = \frac{\Lambda-E_i}{\Lambda-E_0} \ .
\end{eqnarray}
Here, there is no need for a projection since applying
$\Lambda-{\cal H}$ does not give rise to contributions outside the variational
space.  The method converges if
\begin{eqnarray}
\left| \alpha_i \right| < 1 \quad \forall i \ ,
\end{eqnarray}
where $\alpha_i = 0$ would give convergence in one step while ratios
closer to $\pm 1$ yield slower convergence.  This translates into the
condition $\Lambda \ge \Lambda_c=(E_0+E_M)/2$. 
At the critical value $\Lambda_c$, the contribution due to $\Psi_i$ will be
suppressed at the rate
\begin{eqnarray}
\alpha_i = \frac{E_M-E_0+2(E_0-E_i)}{E_M-E_0}
\end {eqnarray}
which becomes smaller as the energy spread in the variational 
subspace $E_M-E_0$ increases.

\section{\label{sec:dmc}Fixed node diffusion Monte Carlo and excited states}

In diffusion quantum Monte Carlo (DMC), the imaginary time evolution operator is
used to stochastically project a trial wave function $\Psi_{\rm T}$ onto the 
lowest energy state $\Phi_0$ of the system to which $\Psi_{\rm T}$ is not 
orthogonal:
\begin{eqnarray}
\label{eq:dmcproj}
\Psi_0 = \lim_{\tau\to\infty} \exp \left\{-\tau{\cal H}\right\}\,\Psi_{\rm T}\,.
\end{eqnarray} 
For fermions, the so-called fixed-node approximation is generally introduced
in order to prevent collapse to the bosonic solution: the result of the 
projection is constrained to have the same nodal surface 
as a given trial wave function $\Psi_{\rm T}$.
This is equivalent to separately solving the Schr\"{o}dinger equation
in the regions of constant sign (nodal pockets) of $\Psi_{\rm T}$ subject to
Dirichlet boundary conditions. The nodal pockets of a ground state wave function 
can be shown to be equivalent, and good ground state trial wave functions
seem to approximate this property well.

In the same way as a trial wave function can be used within the fixed-node 
approximation to prevent collapse to the energetically lower bosonic state, it 
can also be employed for an excited state calculation to prevent collapse to 
lower states. 
If the excited state trial wave function has the exact nodes, the exact excited
state energy is recovered. However, with approximate nodes, the method is not 
variational except for the lowest state of each one-dimensional irreducible 
representation of the pointgroup of the molecule~\cite{needsvari}.


\begin{thebibliography}{999}
\bibitem{non_loc_pseudo} If non-local pseudopotentials are used, the trial wave function
including the Jastrow factor is used to localize the pseudopotential so that the DMC energy 
will depend on the Jastrow factor.
\bibitem{directmin} An alternative could be to minimize $\chi^2$ directly along the
lines of variance minimization. This approach would avoid to compute and to invert the
covariance matrix in Eq.~\ref{eq:lineq_qmc}.
\bibitem{jastrow_linear} We assume that the Jastrow factor does not introduce 
linear dependencies among the variational parameters.
\end{thebibliography}
\end{document}